\documentclass[a4paper,12pt]{article}
\usepackage{amssymb,latexsym}

\makeatletter \@addtoreset{equation}{section} \makeatother

\begin{document}

\begin{titlepage}

\thispagestyle{empty}

\begin{flushright}
\hfill{HU-EP-03/82} \\
\hfill{hep-th/0312210}
\end{flushright}

\vspace{35pt}

\begin{center}{ \LARGE{\bf
$D=4$, ${\cal N}=2$ Gauged Supergravity\\[5mm]
in the Presence of Tensor Multiplets.}}

\vspace{60pt}

{\bf G. Dall'Agata$^\dag$,  R. D'Auria$^\bigstar$, L.
Sommovigo$^\bigstar$ and S. Vaul\'a$^\bigstar $}

\vspace{15pt}

$^\dag${\it Humboldt Universit\"at zu Berlin, Institut f\"ur Physik,\\
Newtonstrasse 15, D-12489 Berlin Adlershof , Germany}\\[1mm] {E-mail:
dallagat@physik.hu-berlin.de}

\vspace{15pt}

$^\bigstar${\it Dipartimento di Fisica, Politecnico di Torino \\
C.so Duca degli Abruzzi, 24, I-10129 Torino, and\\
Istituto Nazionale di Fisica Nucleare, \\
Sezione di Torino,
Italy}\\[1mm] {E-mail: riccardo.dauria@polito.it, \\
luca.sommovigo@polito.it, silvia.vaula@polito.it}

\vspace{50pt}

{ABSTRACT}

\end{center}

\medskip

Using superspace techniques we construct the general theory
describing $D=4$, ${\cal N}=2$ supergravity coupled to an
arbitrary number of vector and scalar--tensor multiplets. The
scalar manifold of the theory is the direct product of a special
K\"ahler and a reduction of a Quaternionic--K\"ahler manifold. We
perform the electric gauging of a subgroup of the isometries of
such manifold as well as ``magnetic'' deformations of the theory
discussing the consistency conditions arising in this process. The
resulting scalar potential is the sum of a symplectic invariant
part (which in some instances can be recast into the standard form of the gauged
${\cal N} =2$ theory)
and of a non--invariant part, 
both giving new deformations. We also show the
relation of such theories to flux compactifications of type II
string theories.

\end{titlepage}

\newpage

\baselineskip 6 mm

\section{Introduction}

Compactifications of type II string theory on Calabi--Yau manifolds
provide effective four--dimensional theories which can be described
by ${\cal N} =2$ supergravity coupled to matter.
The theories  obtained in this way naturally contain
tensor
multiplets, but, using the known duality relation between tensor  and
scalar fields in four dimensions, one can use the standard
formulation in
terms of hypermultiplets.
On the other hand, if in addition to the metric one gives non--trivial expectation
values to the other fields of the ten--dimensional theory, the
effective
supergravity in four dimensions is deformed and
various fields become massive.
Among these there are the tensors which therefore cannot be
dualized into scalars anymore.

These type of deformations are usually described by gauged
supergravity theories and the general couplings for the $d=4$,
${\cal N} =2$ case have been worked out in
\cite{Andrianopoli:1996vr,Andrianopoli:1997cm}. However, the
description provided in
\cite{Andrianopoli:1996vr,Andrianopoli:1997cm} does not include
(massive) tensor multiplets and it is therefore difficult to
establish its relation with flux compactifications of type II
strings. Moreover, it is known that the gauging procedure may give
inequivalent deformations of theories which can be mapped onto
each other before the gauging is performed.
For these reasons it is
quite important to build the general theory describing
four--dimensional ${\cal N} =2$ gauged supergravity coupled to
tensor multiplets and possibly to establish the relation of such a
formulation with the standard one of \cite{Andrianopoli:1997cm}.

Although effective theories of type II compactifications with
fluxes have been described in
\cite{vari,Michelson:1997pn,Taylor:1999ii,
Dall'Agata:2001zh,Louis:2002ny,Andrianopoli:2003jf,Angelantonj:2003zx}
and especially \cite{Louis:2002ny} shows the
explicit appearance of tensor fields, we provide here the general
construction of ${\cal N} =2 $ supergravity with tensor multiplets and
describe its gauging also finding new deformations.
The starting point for our construction\footnote{For earlier work see
\cite{deWit:1983na}.} is given by \cite{Theis:2002er}, where the
tensor multiplet couplings to supergravity were described and some
of the relations on the scalar geometry of the theory were given.

In this paper we improve the results of \cite{Theis:2002er} by
extending the couplings to vector multiplets and by performing the
gauging of the theory. More in detail, by using superspace
techniques we get the general supersymmetry rules after the
dualization of some of the scalars of the hypermultiplets into
tensor fields. As a first step in the construction, we discuss the
necessary conditions that have to be satisfied in order to perform
such a dualization. We then describe the constraints on the
resulting geometry and show their relations with the underlying
quaternionic geometry of the hypermultiplet scalar
$\sigma$--model. In particular, we will show how these constraints
can be understood from a ``Kaluza--Klein'' perspective. When the
quaternionic manifold is given by a homogeneous space, the
dualization procedure yields a space which can be described as the
reduction of the original one by removing some of its nilpotent
generators under the solvable decomposition.

As a next step we perform the gauging of the theory. After
dualization, not all of the isometries of the original manifold
remain  isometries of the final scalar manifold. Moreover, some of
these act non--trivially on the tensor fields and therefore cannot
become local symmetries without leading to non--linear couplings
for the tensor fields. We then discuss which isometries can be
made local and therefore ``gauged''. Always using the superspace
formalism we compute the fermion shifts  which restore the
supersymmetry of the theory and give rise to a potential
satisfying the supersymmetry Ward identities.

The appearance of tensor fields allows to redefine the gauge field
strengths with a shift proportional to the tensor fields
$F^\Lambda \to F^\Lambda + m^{I \Lambda} B_I$ without breaking
supersymmetry, provided we redefine appropriately the fermion
transformation laws.  Here $m^{I \Lambda}$ are real constants
which can be thought as mass parameters for the tensors.
This kind of extension\footnote{Tensor multiplets exist also in 5
dimensions and they can get masses by the same procedure.
However, since they satisfy first order equations of
motion, the gauging of the theory can give additional couplings with the
vectors \cite{Ceresole:2000jd}.} of the theory was first obtained in
\cite{Romans:1986tw} for six--dimensional supergravity,
further extended in \cite{D'Auria:2000ad}
and shown in Calabi--Yau compactification of Type II theories in
\cite{Louis:2002ny}.

Indeed the gauging we perform after dualization of some of the
hypermultiplets scalars is a standard electric gauging, but the
appearance of the mass parameters $m^{I\Lambda}$ in the definition
of the new gauge field strengths implies the existence of extended
solutions. The shifts of the supersymmetry transformations indeed
acquire some extra terms depending on such parameters so that the
gravitino's and hyperino's shifts are symplectic invariants. This
latter can be interpreted also as a ``magnetic" gauging, though its
definition is not related to the appearance of magnetic gauge
fields.
These would lead to the  construction of
\cite{Michelson:1997pn} whose consistency is problematic, as explained
in \cite{Dall'Agata:2001zh,Louis:2002ny}.

The scalar potential of the theory follows as usual from the
square of the fermionic shifts by using a known Ward identity of
$N$--extended gauged supergravities \cite{D'Auria:2001kv}. Being
the square of symplectic invariant quantities, but for a term
coming from the gaugino shift when non--Abelian isometries of the
Special K\"ahler manifold are gauged, the potential shows
symplectic invariance for Abelian gaugings where such gaugino
contribution does not appear. Therefore the potential can be split
in two parts, one that is explicitly symplectic invariant while
the other is not. The first part is in particular the one which
can be obtained by gaugings of translational isometries and it is
therefore directly related to the one which follows from flux
compactifications \cite{Louis:2002ny}. However we can not reduce
by a symplectic transformation this part to the standard one
described in \cite{Andrianopoli:1997cm}, unless we have a single
tensor or in the case where all the symplectic vectors of the
electric and magnetic charges are parallel vectors. Therefore,
except for this particular situation, the magnetic charges give a
genuine deformation of the theory with respect to the standard
$\mathcal{N}=2$ supergravity. This is a fortiori true for non
abelian gaugings, since a non symplectic
invariant extra term is present.

It is interesting to point out
that in this process we also get new consistency conditions with
respect to those of the underlying quaternionic geometry. An
especially interesting condition is given by the requirement that
a certain combination of the  mass parameters must belong to the
center of the gauged Lie Algebra. This condition is certainly
satisfied in our case because each term of the linear combination
is in the center of the algebra since we assume that the
isometries that can be gauged electrically are only those which
commute with the translational isometries of the dualized scalars.
As we will show, these mass parameters can be interpreted as
``magnetic" Killing vectors and therefore this condition has a
natural interpretation as the fact that the electric and magnetic
generators should commute.

\medskip

The plan of the paper is the following. After this introduction,
in section \ref{tens} we review the dualization procedure in order
to fix notations,  discuss the geometry of the scalar manifold and
give the couplings of ${\cal N} = 2$ supergravity to vector,
tensor and hypermultiplets. In section \ref{gauging} we discuss in
detail the gauging procedure. By solving the Bianchi identities of
the various fields using superspace techniques we obtain the
shifts to the supersymmetry transformations and determine the
potential of the coupled theory. We then discuss the properties of
such a potential in section \ref{potential} detailing the
relations with flux compactifications and with the standard theory
of \cite{Andrianopoli:1997cm}. We also give an Appendix where some
explicit examples of dualizations are discussed.

\section{Tensor multiplets coupled to supergravity and vector multiplets}
\label{tens}

As already explained in the introduction, the standard ${\cal N} =
2$ supergravity contains both vector multiplets and
hypermultiplets.
The $\sigma$--model parametrized by the scalars of the
vector multiplets is a special K\"ahler manifold while for the
hypermultiplets the $\sigma$--model is quaternionic--K\"ahler.
When we dualize
some of the scalars of the quaternionic manifold to tensor fields,
the geometry of the hypermultiplets sector is modified and, in
particular, it is no more quaternionic.
As a first step in our
construction we will therefore review the dualization procedure so
that we can also fix our notations.

Before entering the details of the dualization procedure some
comments are in order. Quaternionic--K\"ahler manifolds do not
necessarily admit isometries. On the other hand, the standard
dualization of a hypermultiplet scalar field $q$ into a tensor
$B_{\mu\nu}$ requires that $q$ appears in the Lagrangian only
through its derivatives so that the Lagrangian is invariant under
constant shifts
\begin{equation}
q \to q + \eta\,. \label{eq:shi}
\end{equation}
Moreover, if one wants to obtain more than one tensor field by
this procedure, the isometries associated to the dualized scalars
should commute, so that all of them can be described by equation
(\ref{eq:shi}) at the same time. Notice that so far we did not
make any distinction between compact and non--compact isometries,
though as it is shown in the third example of the appendix, the
resulting physics is very different, actually we get in general a
singular Lagrangian.

\subsection{Dualization of the commuting isometries}

Dualization of the commuting isometries in the quaternionic
manifold spanned by the hypermultiplets can be done in the usual
way by a Legendre transformation on the quaternionic coordinates
$q^{\hat{u}}$, $\hat{u}=1,\dots,4m$ appearing in the Lagrangian
covered by derivatives. Here and in the following, we are using
for the quaternionic geometry the notations given in
\cite{Andrianopoli:1997cm,D'Auria:2001kv}.

Suppose  we partition the coordinates $q^{\hat{u}}$,
$\hat{u}=1,\dots,4m$ into the two subsets $q^u$, $u=1,\dots,n$ and
$q^I$, $I=1,\dots,4m-n$, where $q^I$ are the coordinates we want to
dualize. Since we are interested in the geometry of the resulting
manifold $\mathcal{M}_n$ we just consider the dualization of the
quaternionic kinetic term $\mathcal{L_K}$ in the general
$\mathcal{N}=2$ Lagrangian.
If $d q^I$ is considered as an
independent 1-form field, $d q^I=\Phi^I$ and if we add a $3$--form
Lagrangian multiplier $H_I=d B_I$ to $\mathcal{L_K}$, we have:
\begin{eqnarray}
\label{dua}
&&\mathcal{L_K}= -h_{\hat{u}\hat{v}}\,dq^{\hat{u}}\wedge *
dq^{\hat{v}}\Rightarrow \\ \nonumber
&&-h_{uv}\,dq^{u}\wedge*dq^{v}-2h_{uI}\,dq^{u}
\wedge*\Phi^I-h_{IJ}\,\Phi^I\wedge*\Phi^J+\Phi^I\wedge H_I\,.
\end{eqnarray}
Varying ${\cal L_K}$ with respect to
$B_I$ we find that $\Phi^I$ is closed and therefore one can
equate $\Phi^I= dq^I$.
The variation with respect to $\Phi^I$ gives
\begin{equation}\label{phi}
*\Phi^I= M^{IJ}\left(\frac {1}{2}H^J-h_{uJ}*dq^u\right)
\end{equation}
and substituting this in (\ref{dua}) we find
\begin{equation}
\label{dualLagr}
\mathcal{L_K}^{(dual)}=-g_{uv}dq^u\wedge *dq^v +\frac
{1}{4}M^{IJ}H_I\wedge H_J-M^{IJ}h_{uI}\,dq^u\wedge H_J\,,
\end{equation}
where
\begin{equation}
\label{metric1}
 g_{uv}=  h_{uv}-M_{IJ}A^I_u A^J_v\,, \quad M_{IJ}\equiv
 h_{IJ}\,, \quad M_{IJ}A^J_u=h_{Iu}\,,\quad
\end{equation}
and we have defined $M_{IJ}M^{JK}=\delta^K_I$.

As already observed in \cite{Theis:2002er} the metric for the residual
scalars, the metric for the kinetic term of the tensors and for
their couplings to the scalars, correspond to the Kaluza--Klein
decomposition of the quaternionic metric

\begin{equation}
\label{quatmetr}
h_{\hat u\hat
v}=\epsilon_{AB}\,\mathbb{C}_{\alpha\beta}\,\mathcal{U}_{\hat
u}^{A\alpha}\,\mathcal{U}_{\hat v}^{B\beta}
\end{equation}
that is
\begin{equation}
\label{quatmetric}
h_{\hat u\hat v}
=\begin{pmatrix}{g_{uv}+M_{IJ}A^I_u A^J_v&M_{JK}A^K_u\cr
M_{IK} A^K_v&M_{IJ}}
\end{pmatrix}\,,
\end{equation}
\begin{equation}
\label{inverse} h^{\hat u\hat v}
=\begin{pmatrix}{g^{uv}&-g^{uw}A^J_w \cr
-g^{vw}A^I_w & M^{IJ}+g^{uv} A^I_u A^J_v}
\end{pmatrix}\,.
\end{equation}
The analogous dualization of $\mathcal{L_K}$ made in terms of the
quaternionic vielbein $\mathcal{U}_{\hat{u}}^{A\alpha}$ gives
\begin{equation}
\label{U}
\mathcal{U}_{ u}^{A\alpha}=P^{A\alpha}_{u}+
A^{I}_u\mathcal{U}_{I}^{A\alpha}\,,
\end{equation}
with
\begin{equation}
\label{metric}
P^{A\alpha}_{u}P^{B\beta}_{u}\,{\mathbb C}_{\alpha
\beta}\,\epsilon_{AB}=g_{uv}\,.
\end{equation}

The fact that the reduction of the quaternionic metric has the
same structure as a Kaluza--Klein reduction on the torus was to be
expected since in both cases we consider isometries described by
constant Killing vectors. However, while in the ordinary
Kaluza--Klein the Killing vectors are generators of compact
isometries, here we consider only Killing vectors associated to
non compact isometries, namely translations. Indeed if we perform
a dualization of a compact coordinate covered by the derivative in
a generic $\sigma$-model then, as shown in the example 3 of the
appendix, one usually obtains a singular Lagrangian, while the
dualization of non-compact coordinates gives a regular Lagrangian.
The geometrical procedure associated to such dualization can be
easily described in the case of quaternionic $\sigma$-models which
are symmetric spaces G/H. Indeed let us perform the solvable
decomposition of the Lie algebra G as
\begin{equation}
\label{decomp}
\mathfrak{g}=\mathfrak{h}\oplus
Solv(\mathfrak{g}) = \mathfrak{h} \oplus\mathbb{D}
\oplus\mathbb{N}\,,
\end{equation}
where $\mathbb{D}$ is the non-compact Cartan subalgebra and
$\mathbb{N}$ denote the nilpotent part of the algebra. We consider
in $\mathbb{N}$ the maximal abelian subalgebra of maximal
dimension which can be generated by translational Killing vectors.
Considering any subset of such Killing vectors and deleting the
corresponding generators in $\mathfrak{g}$ corresponds to dualizing
the isometries associated to the coordinates $q^I$. In the
appendix we give two examples of such procedure for the
quaternionic manifolds $SU(2,1)/SU(2) \times U(1)$ and
$SO(1,4)/SO(4)$. Geometrically this corresponds to considering the
quaternionic manifold as a fiber bundle whose base space is
$\mathcal{M}_n$ and fiber given by the set of coordinates
corresponding to the commuting Killing vectors. The projections
along the fibers correspond to our dualization procedure.

\subsection{Reduction of the quaternionic geometry after dualization}
\label{quatern}

It is interesting to reduce the coordinate indices of the
quaternionic   geometry for some important formulae. Let us first
consider the quaternionic relation\footnote{Throughout the paper
we use hats to denote quantities referring to the original
quaternionic--K\"ahler manifold and the same symbols without hats
for the reduced one ${\cal M}_n$.}
\begin{equation}
({\cal U}^{A\alpha}_{\hat{u}} {\cal U}^{B\beta}_{\hat{v}} + {\cal
U}^{A\alpha}_{\hat{v}} {\cal
 U}^{B\beta}_{\hat{u}}) {\mathbb C}_{\alpha\beta}= h_{\hat{u}\hat{v}}
\epsilon^{AB}\,.
\label{piuforte1}
\end{equation}
where  $\epsilon^{AB}$ and ${\mathbb C}_{\alpha\beta}$ are the
usual antisymmetric metrics used to raise and lower indices of
${\rm SU}(2)$ and ${\rm Sp}(2m)$. Splitting (\ref{piuforte1}) on
the $uv\,\,,uI\,\,,IJ$ indices we find three equations which imply
\begin{eqnarray}
\label{rel}
(P^{A\alpha}_u P^{B\beta}_v +
P^{A\alpha}_v P^{B\beta}_u) {\mathbb C}_{\alpha\beta}&=& g_{uv}
\epsilon^{AB} \,,
\label{piuforte2}\\
({\cal U}^{A\alpha}_I {\cal U}^{B\beta}_J + {\cal U}^{A\alpha}_J
{\cal U}^{B\beta}_I) {\mathbb C}_{\alpha\beta}&=& M_{IJ} \epsilon^{AB}\,,
\label{piuforte0}\\
\label{piuforte3} P^{A\alpha}_u {\cal U}_{B\alpha I}+{\cal
U}^{A\alpha}_I P_{B\beta u}&=&0\,.
\end{eqnarray}
Furthermore, using (\ref{inverse}) we can easily obtain
\begin{eqnarray}
\label{invviel1}
{\cal{U}}^{A\alpha u} &=&g^{uv} P^{A\alpha}_v \equiv P^{u A \alpha}\,,\\
{\cal{U}}^{A\alpha I} &=& M^{IJ} {\cal{U}}^{A\alpha}_J - g^{vw}
A^I_w P^{A\alpha}_v \,.\label{invviel2}
\end{eqnarray}

Further insight into the geometrical structure of the sigma model
$\mathcal{M}_n$ is obtained by reduction of the quaternionic
indices $\hat{u}$, $\hat{v}$ for the  connections and curvatures
of the holonomy group contained in ${\rm Sp}(2m)\times {\rm SU}(2)$. The
${\rm SU}(2)$ curvature 2--form is defined as
\begin{equation}
\hat\Omega^x=d\hat\omega^x+\frac{1}{2}\epsilon^{xyz}\hat\omega^y\wedge\hat\omega^z\,,
\end{equation}
where $\hat \omega^x$ is the ${\rm SU}(2)$ connection.
Defining
\begin{equation}
\hat \omega^{AB}=\frac{\rm i}{2}\sigma_x^{AB}\hat \omega^x\,, \quad
\hat \omega^x =-{\rm i}\, \hat \omega^{AB}\sigma^x_{AB}\,,
\end{equation}
where $\sigma_x^{AB}$
are symmetric matrices related to the Pauli matrices
\begin{equation}
\sigma_x^{AB}\equiv\epsilon^{CA}\sigma_{x\,C}^{\ \ B}\,,
\end{equation}
one obtains
\begin{equation}\label{tetto}
\hat\Omega^{AB}\equiv d\hat\omega^{AB}
+\hat\omega^{(AC}\wedge\hat\omega_C^{\phantom{A}B)}\,.
\end{equation}
Reducing the coordinates indices we find for the connections
\begin{eqnarray}
\hat\omega^{AB}_u&=&\hat\omega^{AB}_{C\alpha}\mathcal{U}_u^{C\alpha}=
\hat\omega^{AB}_{C\alpha}\left(P_u^{C\alpha}+A^I_u\mathcal{U}_I^{C\alpha}\right)
\equiv\omega^{AB}_u+A^I_u\omega^{AB}_I\,,
\label{conn1}\\
 \hat\omega^{AB}_I&=&\omega^{AB}_I\,,
\label{conn2}\\
\hat\Delta^{\alpha\beta}_u&=&\hat\Delta^{\alpha\beta}_{C\alpha}\mathcal{U}_u^{C\alpha}=
\hat\Delta^{AB}_{C\alpha}\left(P_u^{C\alpha}+A^I_u\mathcal{U}_I^{C\alpha}\right)
\equiv\Delta^{\alpha\beta}_u+A^I_u\Delta^{\alpha\beta}_I\,,
\label{conn3}\\
\hat\Delta^{\alpha\beta}_I&=&\Delta^{\alpha\beta}_I\,.
\label{conn4}
\end{eqnarray}

It is also important to reduce the ${\rm SU}(2)$ curvature and its
relation to the quaternionic prepotential. Using equations
(\ref{conn1}), (\ref{conn2}) and the fact that after dualization
all the quantities depend just on the $q^u$ so that the derivative
$\partial_I$ is always zero, one obtains from (\ref{tetto})
\begin{eqnarray}
\label{curv1}\hat\Omega^{AB}_{uv}&=&
\Omega^{AB}_{uv}+F^I_{uv}\omega^{AB}_I+2A^I_{[v}\Omega_{u]I}^{AB}+A^I_uA^J_v\Omega^{AB}_{IJ}\,,\\
\hat\Omega^{AB}_{uI}&=&\Omega^{AB}_{uI}+A^J_u\Omega^{AB}_{JI}\,,\\
\label{curv3}\hat\Omega^{AB}_{IJ}&=&\Omega^{AB}_{IJ}\,,
\end{eqnarray}
where the components of the reduced Lie algebra valued ${\rm
SU}(2)$ curvatures are defined as usual:
\begin{eqnarray}
\label{redcurv1}\Omega^{AB}_{uv}&\equiv&
\partial_{[u}\omega_{v]}^{AB}+\omega_{[u}^{AC}\omega_{v]C}^{\phantom{ca}\phantom{ca}B}\,,\\
\Omega^{AB}_{uI}&\equiv&\partial_{[u}\omega_{I]}^{AB}+
\omega_{[u}^{AC}\omega_{I]C}^{\phantom{ca}\phantom{ca}B}=
\frac{1}{2}\nabla_u\omega_I^{AB}\,,\\
\Omega^{AB}_{IJ}&\equiv&\partial_{[I}\omega_{J]}^{AB}+\omega_{[I}^{AC}\omega_{J]C}^{\phantom{ca}\phantom{ca}B}=
\omega_{[I}^{AC}\omega_{J]C}^{\phantom{ca}\phantom{ca}B}\,.
\label{redcurv3}
\end{eqnarray}
On the other hand, for the
quaternionic--K\"ahler geometry the ${\rm SU}(2)$ curvature $\hat\Omega^{AB}$
is defined in terms of the quaternionic vielbeins as
\begin{equation}
\hat\Omega^{AB}=-\mathcal{U}^A{}_\alpha
\wedge\mathcal{U}^{B\beta}\,,
\end{equation}
and therefore reducing the r.h.s. we can also get
\begin{eqnarray}
\label{zurp1}\hat\Omega^{AB}_{uv}&=& -\mathcal{U}_{[u}^A{}_\alpha
\mathcal{U}_{v]}^{B\alpha}=-P_{[u}^A{}_\alpha
P_{v]}^{B\alpha}-A^I_{[u}A^J_{v]}
\mathcal{U}_{[I}^A{}_\alpha \mathcal{U}_{J]}^{B\alpha}\,,\\
\hat\Omega^{AB}_{uI}&=&-\frac{1}{2} \left(
\mathcal{U}_u^A{}_\alpha \mathcal{U}_I^{B\alpha} -
\mathcal{U}_I^A{}_\alpha \mathcal{U}_u^{B\alpha} \right)=
-P_{u\,\alpha}^{(A} \mathcal{U}_I^{B)\alpha} + A^J_u
\Omega^{AB}_{JI} \,,
\\
\hat\Omega^{AB}_{IJ}&=&-\mathcal{U}_{[I}^A{}_\alpha
\mathcal{U}_{J]}^{B\alpha}\,.
\label{zurp3}
\end{eqnarray}
{From} equations (\ref{curv1})--(\ref{curv3}),
(\ref{redcurv1})--(\ref{redcurv3}), (\ref{zurp1})--(\ref{zurp3}) we
readily obtain
\begin{eqnarray}
\Omega^{AB}_{uv}&=&-\nabla_{[u}(A^I_{v]}\omega_I^{AB})-P^A_{\alpha[u}P_{v]}^{B\alpha}\label{ouv}\,,\\
\nabla_u\omega_I^{AB}&=&-2P^{(A}_{u\alpha}\mathcal{U}_I^{B)\alpha}\label{oiv}\,,\\
\omega_{[I}^{AC}\omega_{J]C}^{\phantom{ca}\phantom{ca}B}&=&
-\mathcal{U}^A_{\alpha [I}\mathcal{U}_{J]}^{B\alpha}\,.
\label{oij}
\end{eqnarray}

Furthermore, from the reduction of the quaternionic torsion
equation
\begin{equation}
\hat{\nabla}_{[\hat u}\mathcal{U}^{A\alpha}_{\hat
v]}=0\label{nablu}
\end{equation}
we obtain the following relations
\begin{eqnarray}
\nabla_{[u}P^{A\alpha}_{v]}&=&-F^I_{uv}\mathcal{U}_I^{A\alpha}\,,
\quad\quad
F^I_{uv}\equiv\partial_{[u}A^I_{v]}\,,
\label{tors1}\\
\nabla_{u}\mathcal{U}_I^{A\alpha}&=&
\omega^A_{I\,B}P_u^{B\alpha}+\Delta^\alpha_{I\,\beta}P_u^{A\beta}\,,\\
\omega^A_{\phantom{A}B[I}\mathcal{U}_{J]}^{B\alpha}&=&0\,.
\end{eqnarray}
Note that in equation (\ref{tors1}) there appears a ``torsion"
term $ F^I_{uv}$; nevertheless this term is not related to the
torsion of the connection on the reduced scalar manifold. In fact,
from the torsionless equation of the original quaternionic
manifold (\ref{nablu}) we see that the covariant derivative acting
on the quaternionic vielbein can be split as
\begin{equation}
\hat\nabla_{u}\mathcal{U}^{A\alpha}_v\equiv\partial_u
\mathcal{U}^{A\alpha}_v-\hat\Gamma_{uv}^{\
w}\mathcal{U}^{A\alpha}_w-\hat\Gamma_{uv}^{\
I}\mathcal{U}^{A\alpha}_I+
\hat\omega_{u}{}^{A}{}_B\mathcal{U}^{B\alpha}_v+
\hat \Delta_{u}{}^\alpha{}_\beta\mathcal{U}^{A\beta}_v\label{splitder}\,,
\end{equation}
where the quantities appearing in (\ref{splitder}) were defined in
equations
(\ref{conn1}), (\ref{conn3}), (\ref{U}) and $\hat\Gamma$ are the
components
of the Levi--Civita connection associated to the quaternionic metric
$h_{\hat u\hat v}$.
Considering the antisymmetrizations on the indices
$u,\,v$, the contribution of the Levi--Civita connection vanishes and
substituting the expressions  (\ref{conn1}), (\ref{conn3}), (\ref{U})
one obtains equation (\ref{tors1}).
It is now evident that the
torsion term  $ F^I_{uv}$ arising in the reduction of the
quaternionic--K\"ahler geometry is associated to the torsion of the connection
$\Gamma_{uv}^{\ I}$ which is not the connection on the scalar
manifold ${\cal M}_n$. Indeed from the statement that the
quaternionic metric is covariantly constant
\begin{equation}
\label{cocost1}
\hat\nabla_{\hat u}h_{\hat v\hat w}=0
\end{equation}
one can deduce that also the metric on the reduced scalar manifold
${\cal M}_n$ is covariantly constant
\begin{equation}
\label{cocost2}
\nabla_{ u}g_{ v w}=0
\end{equation}
where $\nabla_u$ is the covariant derivative with respect to the
the Levi--Civita connection on the reduced scalar manifold with
metric $g_{uv}$.

On the other hand, one can observe that the rectangular matrix
$P_u^{A\alpha}$ is not the vielbein on the reduced manifold. Such
vielbein 1--form, which we may denote $E^i$ ($i=1,\dots n$),  must
satisfy $\nabla E^i=0$  where $\nabla$ contains the spin
connection of $\mathcal{M}_n$, in accordance to the fact that the
manifold is torsionless, according to equation (\ref{cocost2}).

{From} equation (\ref{tors1}) it is also possible to evaluate the
Riemannian curvature on the reduced scalar manifold acting with a
further covariant derivative; one obtains:
\begin{equation}
\mathcal{R}_{vzwu}=-\Omega^{AB}_{uv}P_{z\,A\alpha}P^\alpha_{v\,B}-\mathbb{R}^{\
\alpha}_{wu\,\beta}P_{z\,A\alpha}P_v^{A\beta}-M_{IJ}F^I_{v[u}F^J_{w]z}\,.
\label{riemcur}
\end{equation}
which shows explicitly, as already observed in reference
\cite{Theis:2002er}, that the holonomy of the reduced scalar
manifold is not contained in ${\rm SU}(2)\times {\rm Sp}(2m)$.

Finally, for the interpretation of the gauged supergravity theory
coupled to tensor multiplets, it is important to reduce the
fundamental relation defining the quaternionic prepotential (see for
instance \cite{Andrianopoli:1997cm,D'Auria:2001kv}), namely
\begin{equation} \label{prepotente}
2k_\Lambda^{\hat{u}}\hat\Omega^x_{\hat u\hat v}=-\nabla_{\hat
v}\mathcal{P}^x_\Lambda\,.
\end{equation}
Writing equation
(\ref{prepotente}) with explicit free index $\hat v\rightarrow (v,\,
I)$ and setting $k^I_\Lambda=0$ we get two equations which combined
give the following relations
\begin{eqnarray}
2k^u_\Lambda(\Omega^x_{uv}+
F^I_{uv}\omega^x_I)-k^u_\Lambda
A^I_u\nabla_v\omega^x_I&=&-\nabla_v\mathcal{P}^x_\Lambda\,,\\
k^u_\Lambda(\nabla_u\omega^x_I+2A^J_u\Omega^x_{JI})+
\epsilon^{xyz}\omega^y_I\mathcal{P}^z_\Lambda&=&0\,.
\label{prepoz}
\end{eqnarray}

\subsection{Coupling of the scalar--tensor multiplet to $\mathcal{N}=2$\\ ungauged supergravity with vector multiplets}

After the dualization of some of the hypermultiplets, the original
quaternionic--K\"ahler manifold reduces to a scalar manifold
$\mathcal{M}_n$ whose scalar fields are part of the so called
scalar--tensor multiplet
\begin{equation}
\left\{B_{I\,\mu\nu},\,\zeta^\alpha,\,\zeta_\alpha,\,q^u\right\}\,,
\end{equation} where
the index $I$ takes $n_T$ values, $n_T$ being the number of
coordinates covered by derivative which have been dualized,
$\alpha=1,\dots m$ and $u=1,\dots n$, $n=4m-n_T$.

We recall \cite{Andrianopoli:1997cm} that the content of the gravitational and vector
multiplets is
\begin{eqnarray}
&\left\{V^a_\mu,\, A^0_\mu,\,
\psi_A,\,\psi^A\right\}\,,&\\
&\left\{A^i_\mu,\,\lambda^{iA},\,\lambda^{\bar{\imath}}_A,\,
z^i\right\}\,,&
\end{eqnarray}
where $i=1,\dots n_v$, $n_v$ being the number of
vector multiplets. The lower ${\rm SU}(2)$ index $A$ on the gravitino
$(\psi_A,\,\psi^A)$ corresponds to positive chirality, while the
corresponding upper index denotes negative chirality.
The opposite convention is adopted for the gauginos
$(\lambda^{iA},\,\lambda^{\bar{\imath}}_A)$.
Finally, for the hyperinos fields $(\zeta_\alpha,\,\zeta^\alpha)$,
lower and upper indices correspond to positive and negative chirality
respectively.

In order to couple the scalar--tensor multiplet to the
$\mathcal{N}=2$ supergravity in the presence of vector multiplets,
we use the superspace approach writing the curvatures of the
various fields in superspace and solving the corresponding Bianchi
identities. The curvatures are defined as follows
\cite{Andrianopoli:1997cm,D'Auria:2001kv}
\begin{eqnarray}
T^a & \equiv & {\cal D}V^a- {\rm i} \, \bar\psi_A\wedge
\gamma^a\psi^A=0\,,
\label{torsdef}\\
\rho_A &  \equiv & d\psi_A-{1\over 4} \gamma_{ab} \,
\omega^{ab}\wedge\psi_A+ {{\rm i} \over 2} {\cal Q}\wedge \psi_A +
\omega_A^{~B}\wedge \psi_B
\equiv \nabla \psi_A\,,
\label{gravdefdown} \\
\rho^A & \equiv & d\psi^A-{1\over 4} \gamma_{ab} \,
\omega^{ab}\wedge\psi^A -{{\rm i} \over 2} {\cal Q}\wedge\psi^A
+\omega^{A}_{\phantom{A}B} \wedge \psi^B \equiv \nabla \psi^A\,,
\label{gravdefup} \\
R^{ab} & \equiv & d\omega^{ab}-\omega^a_{\phantom{a}c}\wedge
\omega^{cb}\,.
\label{riecurv}
\end{eqnarray}
In the vector multiplet sector the curvatures and covariant
derivatives are:
\begin{eqnarray}
\nabla z^i &=& dz^i\,,
\label{zcurv}\\
\nabla {\bar z}^{\bar{\imath}} &=& d{\bar z}^{\bar{\imath}}\,,
\label{zcurvb}\\
\nabla\lambda^{iA} &\equiv & d\lambda^{iA}-{1\over 4} \gamma_{ab}
\, \omega^{ab} \lambda^{iA} -{{\rm i} \over 2} {\cal
Q}\lambda^{iA}+ \Gamma^i_{\phantom{i}j}\lambda^{jA}+
\omega^{A}_{~B} \wedge \lambda^{iB}\,,
\label{lamcurv} \\
\nabla\lambda^{\bar{\imath}}_A &\equiv
&d\lambda^{\bar{\imath}}_A-{1\over 4} \gamma_{ab} \,
\omega^{ab}\lambda^{\bar{\imath}}_A+{{\rm i} \over 2} {\cal
Q}\lambda^{\bar{\imath}}_A+ \Gamma^{\bar{\imath}}_{\phantom{\bar
{\imath}}{\bar{\jmath}}}\lambda^{\bar{\jmath}}_A + \omega_{A}^{~B}
\wedge \lambda^{{\bar{\imath}}}_B\,,
\label{lamcurvb} \\[2mm]
F^\Lambda &\equiv & dA^\Lambda+\, \bar L^\Lambda
\bar\psi_A\wedge\psi_B \epsilon^{AB}+L^\Lambda\bar\psi^A\wedge
\psi^B\epsilon_{AB}\,. \label{Fcurv}
\end{eqnarray}
In the scalar--tensor multiplet sector they are
\begin{eqnarray}
H_I &=& dB_I + 2\, L{}_{I\,C}{}^A \, \bar \psi_A \gamma^a \psi^C V_a \,,\\
\nabla \zeta_\alpha & = & d\zeta_{\alpha} - \frac{1}{4}
\omega^{ab} \gamma_{ab} \zeta_\alpha - \frac{\rm i}{2} \,{\cal Q}
\,\zeta_\alpha +
\Delta_\alpha^\beta \zeta_\beta\,, \\
\nabla \zeta^\alpha & = & d\zeta^{\alpha} - \frac{1}{4}
\omega^{ab} \gamma_{ab} \zeta^\alpha- \frac{\rm i}{2}{\cal Q}
\zeta^\alpha + \Delta^\alpha_\beta \,\zeta^\beta\,,\\
P^{A\alpha}&=&P^{A\alpha}_u dq^u\,,
\label{Pcurv}
\end{eqnarray}
where $\mathcal{Q}$ is the $\rm U(1)$ K\"ahler connection \cite{Andrianopoli:1997cm},
$\Gamma^i_j$ is the Christoffel connection one--form for the special
K\"ahler manifold, $L^\Lambda$ is part of the symplectic section
$(L^\Lambda,\,M_\Lambda)$ of the special manifold, $\omega^{AB}$ and
$\Delta^{\alpha\beta}$
are respectively the ${\rm SU}(2)$ and ${\rm Sp}(2m,\mathbb{R})$
connections (\ref{conn1}), (\ref{conn3}) on the $\mathcal{M}_n$ manifold
and $L_{IC}^{\ A}(q)$ is a tensor to be determined by the Bianchi
identities.

Differentiating equations (\ref{zcurv})--(\ref{Pcurv}) one obtains
the Bianchi identities, whose superspace solutions are given by
the following parameterizations (up to three fermion terms):
\begin{eqnarray}
T^a & = & 0 \,,
\label{torsion}\\
\rho_A & = & V^a V^b \rho_{ab A}+\epsilon_{AB} \,T^-_{ab}
\,\gamma^b \psi^{B}
V^a-\frac{1}{2}h_{Ia}L_{A}^{I\phantom{}B}\psi_BV^a\,,
\label{gravitino}\\[2mm]
F^\Lambda & = & V^a V^b \, F_{ab}^\Lambda + \left({\rm i} f_i^\Lambda\,
\bar \lambda^{iA} \gamma_a \psi^B \epsilon_{AB} + {\rm i} \bar
f^\Lambda_{\bar \imath}\, \bar \lambda^{\bar \imath}_A \gamma_a
\psi_B
\epsilon^{AB}\right) V^a \,, \label{gaugevec}\\[2mm]
\nabla \lambda^{iA} & = & V^a \nabla_a \lambda^{iA} + {\rm i} \,Z^i_a
\,\gamma^a \psi^A + G^{-i}_{ab}\,\gamma^{ab}\psi_{B}\epsilon^{AB}\,,
\label{gaugino}\\[2mm]
H_I&=& H_{I\,abc}V^a V^b V^c + \frac{\rm i}{2}\left(\bar \psi_A
\gamma^{ab} \zeta_\alpha\, g_I^{A\alpha} \, V_a V_b - \bar \psi^A
\gamma^{ab} \zeta^\alpha \, g_{I\,A\alpha}\, V_a V_b \right)\,,
\label{tensor}\\[2mm]
\nabla \zeta_\alpha & =& V^a \nabla_a \zeta_\alpha +{\rm i} P_{aA\alpha}
\gamma^a \psi^A +{\rm i}\,h_{Ia}\, g^I_{A\alpha}\, \gamma^a \psi^A\,,
\label{hyperino}\\[2mm]
\nabla z^i & = & Z^i_a V^a + \bar\psi_{A} \lambda^{iA} \,,
\label{vecscalar}\\[2mm]
P^{A\alpha} & = &P^{A\alpha}_a V^a +\bar\psi^A \zeta^\alpha
+\epsilon^{AB}\mathbb{C}^{\alpha\beta} \bar\psi_B \zeta_\beta\,. \label{hyperscalar}
\end{eqnarray}
We observe that the solution (\ref{torsion})--(\ref{gaugino}) of the
Bianchi identities is the same as the ordinary $\mathcal{N}=2$
supergravity coupled to vector multiplets, except for the extra
term $h_{Ia}L_{A}^{I\phantom{}B}\psi_BV^a$ appearing in
(\ref{gravitino}), where $h_{I\,a}=\epsilon_{abcd}H_I^{bcd}$. In
particular, we have used the well known definitions for the
self--dual dressed vector field strengths:
\begin{eqnarray}
T^{-}_{ab} & = & \left({\cal N}- \bar{\cal
N}\right)_{\Lambda\Sigma}
L^\Sigma F^{-\Lambda}_{ab} \,,\label{fs1}\\
G^{i-}_{ab} & = & \frac{\rm i}{2}g^{i\bar \jmath}f_{\bar
\jmath}^\Gamma \left({\cal N}- \bar{\cal N}\right)_{\Gamma\Lambda}
F^{\Lambda-}_{ab}\,.\label{fs2}
\end{eqnarray}
and
\begin{equation}
f^\Lambda_i\equiv\nabla_i L^\Lambda\,,
\end{equation}
where $\nabla_i$ is the K\"ahler covariant derivative.
Furthermore the Bianchi identities
imply that the tensors $g_I^{A\alpha}$, $P_a^{A\alpha}\equiv
P_u^{A\alpha}\partial_a q^u$ and $L^I_{AB}$ satisfy the following set
of constraints
\begin{eqnarray}
P^u_{A\alpha}\,P_v^{B\alpha}+P^{uB\alpha}\,P_{vA\alpha}&=&\delta^u_v\,\delta^{B}_A\,,
\label{eq:rel4}\\[2mm]
g^{AI\alpha}\,g_{JB\alpha} + g^I_{B\alpha} \,g_J^{A\alpha}&=&\delta^I_J
\,\delta^A_B\,,
\label{eq:rel3}\\[2mm]
P_{aC\alpha} \,g^{IA\alpha}+ P^{A\alpha}_a\, g^I_{C\alpha}&=&0\,,
\label{eq:rel2}\\[2mm]
P_{aC\alpha} \,g^{A\alpha}_I+ P^{A\alpha}_a \,g_{IC\alpha}&=&0\,,
\label{eq:rel2.1}\\[2mm]
\nabla_{a} L_{I}^{CA}  = P_{a\alpha}^{(C}g^{A)\alpha}_I\,,
\label{eq:rel1}\\[2mm]
g_J^{\alpha(A}\,
g_{I\alpha}^{B)}={\rm i}\epsilon_{xyz}\,L_I^y\,L_J^z\,\sigma^{x\phantom{B}A}_B\,.
\label{eq:rel3.1}
\end{eqnarray}
It is important to note that the previous constraints on the
tensors appearing in the scalar--tensor multiplet (coinciding
with those found in reference \cite{Theis:2002er}) are in complete
agreement with the results obtained from the reduction of the
quaternionic geometry of section \ref{quatern}.
Indeed, equation (\ref{eq:rel4}) coincides with equations (\ref{piuforte2})
and equation (\ref{eq:rel2.1}) coincides with
(\ref{piuforte3}) provided we identify
\begin{equation}\label{ug}
g_I^{A\alpha}\equiv\mathcal{U}_I^{A\alpha}\,.
\end{equation}
For what concerns (\ref{eq:rel2}), we can also see that it can also be
recast into  the form of equation (\ref{piuforte3}) upon identifying
\begin{equation}
g^{I\,A\alpha}\equiv
M^{IJ}g^{A\alpha}_J=\mathcal{U}^{I\,A\alpha}+g^{uv}A_u^I
P_v^{A\alpha}\label{gu}
\end{equation}
With the identifications (\ref{ug}), (\ref{gu}) we also obtain (\ref{piuforte0}) from
equation (\ref{eq:rel3}).
Equations (\ref{eq:rel1}), (\ref{eq:rel3.1}) instead coincide with
(\ref{oiv}) and (\ref{oij}) provided we identify
\begin{equation}
L_I^{AB}=-\frac{1}{2}\omega_I^{AB}\,.
\label{identif}
\end{equation}

\medskip

Finally, for the benefit of the reader we write down the
supersymmetry transformation laws of the fields on space--time as
they follow immediately from the Bianchi identities solution
(\ref{torsion})--(\ref{hyperscalar}):
\begin{eqnarray}
\delta\,\psi _{A \mu} &=& {\cal D}_{\mu}\,\varepsilon _A\,
 +\epsilon_{AB} T^-_{\mu \nu}\gamma^{\nu}\varepsilon^B
 \label{trasfgrav0} \\[2mm]
\delta \,\lambda^{iA}&=&
 {\rm i}\,\nabla _ {\mu}\, z^i
\gamma^{\mu} \varepsilon^A +\epsilon^{AB}G^{-i}_{\mu \nu} \gamma^{\mu
\nu} \varepsilon _B \,
\label{gaugintrasfm0}\\[2mm]
 \delta\,\zeta _{\alpha}&=& {\rm i}\,
P_{u\, A\alpha}\, \partial _{\mu}\,q^u \,\gamma^{\mu}
\varepsilon^A+{\rm i} h_{\mu I}g^I_{A\alpha}\gamma^\mu\varepsilon^A
 \, \label{iperintrasf0}\\[2mm]
\delta\,V^a_{\mu}&=& -{\rm i}\,\bar {\psi}_{A
\mu}\,\gamma^a\,\varepsilon^A -{\rm i}\,\bar {\psi}^A _
\mu\,\gamma^a\,\varepsilon_A\\[2mm]
\delta \,A^\Lambda _{\mu}&=& 2 \bar L^\Lambda \bar \psi _{A\mu}
\varepsilon _B
\epsilon^{AB}\,+\,2L^\Lambda\bar\psi^A_{\mu}\varepsilon^B \epsilon_{AB}
+{\rm i} \,f^{\Lambda}_i \,\bar {\lambda}^{iA} \gamma _{\mu}
\varepsilon^B \,\epsilon _{AB} +{\rm i} \, {\bar
f}^{\Lambda}_{{i}^\star} \,\bar\lambda^{{i}^\star}_A
\gamma _{\mu} \varepsilon_B \,\epsilon^{AB} \label{gaugtrasf0}\\[2mm]
\delta B_{\mu\nu\,I}&=&\frac{\rm i}{2}\left(\bar \varepsilon_A
\gamma_{\mu\nu} \zeta_\alpha\, g_I^{A\alpha} - \bar \varepsilon^A
\gamma_{\mu\nu} \zeta^\alpha \, g_{I\,A\alpha}\right)+2 L_{IC}^{\
A}\bar\varepsilon_A\gamma_{[\mu}\psi^C_{\nu]}+2 L_{IC}^{\
A}\bar\psi_{[\mu A}\gamma_{\nu]}\varepsilon^C\\[2mm]
\delta\,z^i &=& \bar{\lambda}^{iA}\varepsilon_A \label{ztrasf0}\\[2mm]
\delta\,z^{{i}^\star}&=& \bar{\lambda}^{{i}^\star}_A \varepsilon^A
\label{ztrasfb0}\\[2mm]
  \delta\,q^u &=& P^u_{\alpha A} \left(\bar {\zeta}^{\alpha}
  \varepsilon^A + C^{\alpha  \beta}\epsilon^{AB}\bar {\zeta}_{\beta}
  \varepsilon _B \right)
\end{eqnarray}

\section{The gauging}
\label{gauging}

{From} the supersymmetry rules described in the previous section we
have seen that the vector multiplet scalar manifold and its
couplings have not been touched by the dualization procedure. This
means that if the scalar manifold of the vector multiplets admits
isometries, these can be gauged without any further restriction
both for abelian and non--abelian gauge groups. On the other hand
we have seen that the hypermultiplet scalar manifold now has been
reduced and the resulting manifold contains the scalars $q^u$ of
the original  hypermultiplets that have not been dualized. This
means that care is needed in order to decide which isometries of
the original quaternionic manifold can be gauged.

On general grounds, before the dualization procedure one could
have,  besides translations, isometries of the scalar manifolds
acting non--trivially on the quaternions which we want to dualize
into tensor fields. These isometries now become symmetries of the
resulting manifold under which the tensor fields are charged. As a
consequence one cannot make these symmetries local without paying
the price of introducing non--linear couplings of the tensor
fields themselves \cite{Ogievetsky:1967ij}. In
what follows we will therefore limit ourselves to the gauging of
symmetries which commute with the translations of the dualized
scalars. One should also remember that in order to gauge
non--abelian groups, these symmetries should also be symmetries of
the special K\"ahler manifold \cite{Andrianopoli:1997cm}.

For these symmetries the gauging procedure is the standard one
outlined in \cite{Andrianopoli:1997cm}.
One adds to the (composite) connections appearing in the
transformation laws of the charged fields
the vector fields (which make this symmetry local) dressed with some
function of the scalar fields.
To be specific, the connections of the vector line bundle ${\cal Q}$,
the connection of the vector tangent bundle $\Gamma^i_j$, the $SU(2)$
connection $\omega^x$ and the  $Sp(2m)$ connection
$\Delta^{\alpha\beta}$ are shifted as
\begin{eqnarray}
\Gamma^i_j & \to & \Gamma^i_j + g \, A^\Lambda \, \partial_j k^i_\Lambda\,,
\label{eq:tangentbundle}\\
{\cal Q} & \to & {\cal Q} + g \, A^\Lambda {\cal P}_\Lambda^0\,,
\label{eq:linebundle}\\
\omega^x & \to & \omega^x + g\, A^\Lambda {\cal P}_\Lambda^x\,,
\label{eq:SU2conn}\\
\Delta^{\alpha\beta} & \to &  \Delta^{\alpha\beta}+g \, A^\Lambda
\partial_uk^v_\Lambda \, {\cal U}^{u\alpha A}{\cal
U}_{vA}^{\beta}\,.\label{eq:SPmbundle}
\end{eqnarray}
These modifications obviously break the supersymmetry of the
Lagrangian which can be restored by adding appropriate shifts to
the supersymmetry rules of the Fermi fields as well as adding mass
terms for the fermions in the Lagrangian at first order in $g$ and
a scalar potential which is of order $g^2$. Moreover the
definitions (\ref{zcurv}), (\ref{zcurvb}), (\ref{Fcurv}) and
(\ref{Pcurv}) get modified to
\begin{eqnarray}
\nabla z^i &=& dz^i +g\, A^\Lambda k_\Lambda^i\,,
\label{zcurvnn}\\
\nabla {\bar z}^{\bar{\imath}} &=& d{\bar z}^{\bar{\imath}}+ g\,A^\Lambda
k_\Lambda^{\bar \imath} \,,
\label{zcurvbnn}\\
F^\Lambda &\equiv & dA^\Lambda+\frac12
\,g\,f^{\Lambda}_{\Gamma\Sigma} \,A^\Gamma \wedge A^\Sigma+ \bar
L^\Lambda \bar\psi_A\wedge\psi_B
\epsilon^{AB}+L^\Lambda\bar\psi^A\wedge \psi^B\epsilon_{AB}\,,
\label{Fcurvnn}\\
P^{A\alpha}&=&P^{A\alpha}_u \left(dq^u + g\,A^\Lambda
\,k^u_\Lambda\right)\,.
\label{Pcurvnn}
\end{eqnarray}

In addition to this standard electric gauging, for which the only
constraint is the choice of generators commuting with the
translations of the dualized scalars $q^I$, we can also introduce
here mass terms for the tensor fields which will enter in the
theory in a way that looks like a ``magnetic gauging". We stress
however, that we are not gauging the magnetic potentials and
therefore the mass parameters can not be identified with magnetic
charges since they are not related to the isometries of the
reduced scalar manifold.

The appearance of tensor fields in the ungauged theory
allows us for the introduction of explicit mass terms for these
fields by redefining the (\ref{Fcurvnn}) curvatures
\begin{equation}
\tilde F^\Lambda_{ab} = F^\Lambda_{ab} + m^{I\Lambda} B_{I ab}\,,\label{Bmasses}
\end{equation}
where $m^{I\Lambda}$ are real constants. By doing so, the tensor
fields appear naked in the Lagrangian and therefore cannot be
trivially dualized back into scalar fields. Moreover this process
will introduce explicit mass terms for $B_I$ and indeed it can be
viewed as the usual St\"uckelberg mechanism, where combinations of
the gauge field strengths are absorbed in the definition of
massive tensor fields. The introduction of these extra terms also
breaks the supersymmetry of the Lagrangian. We will see in a while
that we can again restore it by further shifting the
supersymmetry transformations of the Fermi fields by terms of the
first order in $m$ and by modifying the scalar potential. It is
actually interesting that these shifts will not simply produce
$m^2$ terms in the potential but will also interfere with the
electric shifts giving terms proportional to both couplings.

In the presence of the electric gauging and the mass parameters
$m^{I\Lambda}$ the supersymmetry transformation laws of the fermionic
fields (as they come from the closure of the gauged Bianchi
identities) become:
\begin{eqnarray}
\delta\,\psi _{A \mu} &=& {\cal D}_{\mu}\,\varepsilon _A\,
 +\epsilon_{AB} T^-_{\mu \nu}\varepsilon^B-\frac{1}{2}h_{\mu I}L^{I\,B}_A\varepsilon_B
+{\rm i}\,S_{AB}
\gamma^{\mu}\varepsilon^B\,,
 \label{trasfgrav} \\[2mm]
\delta \,\lambda^{iA}&=&
 {\rm i}\,\nabla _ {\mu}\, z^i
\gamma^{\mu} \varepsilon^A +G^{-i}_{\mu \nu} \gamma^{\mu \nu}
\epsilon^{AB}\varepsilon _B \,+\, W^{iAB}\varepsilon _B\,,
\label{gaugintrasfm}\\[2mm]
 \delta\,\zeta _{\alpha}&=&{\rm i}\, P_{\mu A\alpha}
\gamma^\mu \varepsilon^A +{\rm i}\,h_{I\mu}\, g^I_{A\alpha}\,
\gamma^\mu \varepsilon^A \,+\,N_{\alpha}^A\,\varepsilon_A\,,
\label{iperintrasf}
\end{eqnarray}
where the extra fermionic terms are:
\begin{eqnarray}
S_{AB} & =& \frac{\rm i}{2}\,\sigma^x_{AB}\left(g{\cal
P}^x_\Lambda L^\Lambda - L_{I}^x m^{I\Lambda} M_\Lambda\right)\,,
\label{shiftS}\\[2mm]
N^\alpha_A & = &-2\,g\,\mathcal{U}^\alpha_{Au}k^u_\Lambda L^\Lambda
+g_{IA}^\alpha m^{I\Lambda}M_{\Lambda}\,,
\label{shiftN}\\[2mm]
W^{iAB} & = & -{\rm i}\,g\, \epsilon^{AB} g^{i\bar \jmath}
P_\Lambda f^\Lambda_{\bar \jmath} + {\rm i}\,g^{i\bar \jmath}
\sigma^{AB}_x\left(g{\cal P}^x_\Lambda\bar f_{\bar \jmath}^\Lambda
- L_I^x m^{I\Lambda} h_{\Lambda \bar \jmath}\right)\,,
\label{shiftW}
\end{eqnarray}
where now the self--dual dressed vector field strengths (\ref{fs1}),
(\ref{fs2}) appear combined with the tensor fields as in
(\ref{Bmasses}):
\begin{eqnarray}
T^{-}_{ab} & = & \left({\cal N}- \bar{\cal N}\right)_{\Lambda\Sigma}
L^\Sigma \tilde F^{-\Lambda}_{ab}\,, \\
G^{i-}_{ab} & = & \frac{\rm i}{2}g^{i\bar \jmath} \bar f_{\bar
\jmath}^\Gamma \left({\cal N}- \bar{\cal N}\right)_{\Gamma\Lambda}
\tilde F^{\Lambda-}_{ab}\,.
\end{eqnarray}
As expected, the dual sections of the vector scalar manifold
appear in the magnetic parts of the above expressions
\begin{equation}
M_\Lambda = {\cal N}_{\Lambda\Sigma} L^\Sigma\,,\quad h_{\Lambda i} =
\overline{\cal N}_{\Lambda\Sigma} f^\Sigma_i\,.
\label{eq:deffo}
\end{equation}
We notice that the electric part of the shifts has not changed
from the standard one obtainable in the absence of tensor
multiplets. In (\ref{shiftN}) there appear the Killing vectors of
the isometries gauged of the scalar--tensor $\sigma$--model, which
are the reduction of the isometries of the original
quaternionic--K\"ahler manifold. The prepotentials associated to
these isometries appear in (\ref{shiftS}). The geometrical
interpretation of ${\cal P}_\Lambda^x$ cannot be the same of the
standard theory, but the conditions following from the requirement
of supersymmetry match (\ref{prepoz}), which are the relations
following from the reduction of those existing between
quaternionic Killing vectors and their prepotentials.

Consistency imposes a non--trivial constraint for
the (\ref{shiftS}) and (\ref{shiftN}) shifts
\begin{eqnarray}
4 L_{IC}{}^{(A} \bar S^{B)C}& = &- N^{(B}_{\alpha} g^{A)\alpha}_I\,,\label{NS}\\
4 L_{I(A}{}^{C} S_{B)C} & = &  N^\alpha_{(A} g_{B)\alpha
I}\label{SN}\,.
\end{eqnarray}
The electric part of these equations can be interpreted as the
reduction of the geometric relations between quaternionic Killing
vectors and prepotentials (\ref{prepotente}) on the external
directions $I$, therefore matching (\ref{prepoz}). This can also
be seen as a standard gradient flow equation
\cite{D'Auria:2001kv}. The terms in the shifts proportional to
$m^{I\Lambda}$ entering in equations (\ref{NS}), (\ref{SN})
identically solve them by using (\ref{oij}), where the identification
(\ref{identif}) of the $\omega_I^x$ connection with $L_{I}^x$ is
done.

This last identity can also be interpreted as a gradient flow
equation assuming that the $m^{I\Lambda}$ constants have the
r$\hat{\rm o}$le of ``magnetic" Killing vectors. In this way one
can introduce ``magnetic prepotentials'' ${\cal Q}^{\Lambda\,x}$.
Considering equation (\ref{prepotente}) for dual Killing vectors
$\tilde k^{I\Lambda}$, for which we set $\tilde k^{u\Lambda}=0$,
one obtains
\begin{equation}
2\,\tilde k^{I\Lambda}\Omega^{x}_{IJ}\equiv-\nabla_{J}\mathcal{Q}^{x\Lambda}
\equiv -\epsilon^{xyz} \omega_{J}^y {\cal Q}^{z\Lambda}\,,
\label{eq:Qprep}
\end{equation}
where it was used that ${\cal Q}^{x\Lambda}$ should be independent
on the scalars we dualized.
We can now see that (\ref{NS}) and (\ref{SN}) are the same as
(\ref{eq:Qprep}) if we relate the Killing vector and the
mass parameter as
\begin{equation}
\tilde k^{I\Lambda}=-\frac{1}{2}m^{I\Lambda}\,.
\end{equation}
and define ${\cal Q}^{x\Lambda}$ as
\begin{equation}
{\cal Q}^{x~\Lambda} = -\frac{1}{2} \omega_I^x m^{I\Lambda} = L^x_I
m^{I\Lambda}.
\label{Qdef}
\end{equation}
By doing so almost all the terms in (\ref{shiftS})--(\ref{shiftW})
can be rewritten in terms of
symplectic invariants quantities.
One can indeed introduce the symplectic vector
\begin{equation}
{\cal T}^x= \left\{ {\cal Q}^{x\,\Lambda}\,,g \, {\cal P}^x_{\Lambda}
\right\}
\label{sympllvect}
\end{equation}
and rewrite (\ref{shiftS})--(\ref{shiftW}) as contractions of this
with the other symplectic vectors given by $V \equiv \{L^\Lambda,
M_\Lambda\}$ and their derivatives
\begin{equation}
U_i = \nabla_i V\equiv
\{f^\Lambda_i,\,h_{\Lambda i}\}\,.
\end{equation}
Introducing also the symplectic vector
\begin{equation}
{\cal Z}_A^\alpha = \left\{-g_{IA}^\alpha \,\tilde
k^{I\Lambda}\,,\; g\, k^u_\Lambda
\,\mathcal{U}^\alpha_{uA}\right\}=
 \left\{1/2 \, g_{IA}^\alpha
\,m^{I\Lambda}\,,\;g\, k^u_\Lambda
\,\mathcal{U}^\alpha_{uA}\right\}\,, \label{eq:gradPQ}
\end{equation}
the shifts read
\begin{eqnarray}
S_{AB} & =& \frac{\rm i}{2} \sigma^x_{AB} <V, {\cal
T}^x>\,, \\[2mm]
N^\alpha_A & = &-2\,<V,{\cal Z}^\alpha_A>\,, \\[2mm]
W^{iAB} & = & -{\rm i}\,g\, \epsilon^{AB} g^{i\bar \jmath}
P_\Lambda f^\Lambda_{\bar \jmath} + {\rm i}\,  g^{i\bar \jmath}
\sigma_x^{AB} < U_{\bar\jmath},{\cal T}^x >\,,
\end{eqnarray}
where $<,>$ denotes the symplectic scalar product defined as
follows:
\begin{equation}
\begin{pmatrix}{a^\Lambda &b_\Lambda}\end{pmatrix}
\begin{pmatrix}{0&\delta_\Lambda{}^\Sigma\cr
-{\delta^\Lambda{}_\Sigma}&0}\end{pmatrix}
\begin{pmatrix}{c^\Sigma\cr d_\Sigma}\end{pmatrix}
\end{equation}

This rewriting is especially useful in view of the construction of the
scalar potential.
Since the above expressions are all symplectic invariants, but for the
antisymmetric piece in the gaugino's shift $W^{[AB]}$, also the
corresponding contributions to the scalar potential will have the same
properties.

So far we did not find any further constraint on the possible
gauge group by adding magnetic charges, but we will see that this
will not be the case anymore when considering the supersymmetry
Ward identity of the scalar potential.

\section{The potential}
\label{potential}

The scalar potential can be determined by a general Ward identity
\cite{D'Auria:2001kv} of extended supergravities, which shows that it
follows from squaring the fermion shifts.
In the present case such identity reads
\begin{equation}
\label{ward}
\delta_B^A {\cal V}  \,= \,-12\,\overline{S}^{CA} \, S_{CB}\,+\,g_{i\bar
\jmath}W^{iCA}\,W_{CB}^{\bar \jmath}\,
+\,2N^A_\alpha \, N^\alpha_B\,,
\end{equation}
with $S_{AB}$, $N^\alpha_A$ and $W^{AB}$ given by (\ref{shiftS}), (\ref{shiftN}) and
(\ref{shiftW}) respectively.

As it is clear from the definition of the fermionic shifts, the
right hand side of (\ref{ward}) contains both pieces proportional
to $\delta^A_B$ as well as to $(\sigma^x)^A_B$. In order for the
theory to be supersymmetric one has to prove that the parts which
are $SU(2)$ Lie Algebra valued vanish identically. Since the
various shifts contain expressions proportional to the electric
coupling constant and others proportional to the ``magnetic" one,
their squares have three type of pieces which have to be set to
zero separately. The first condition follows from the terms
proportional to $g^2$ and it is the simple reduction of the
quaternionic identity
\begin{equation}
\Omega_{uv}^x \, k^u_\Lambda k^v_\Sigma - \frac12 \, \epsilon^{xyz} {\cal
P}^y_\Lambda {\cal P}^z_\Sigma + \frac12 \, f^\Delta_{\Lambda\Sigma}
\, {\cal P}^x_\Delta = 0\,.
\label{eq:quatid}
\end{equation}
Also the $m^2$ piece is identically satisfied. This can be seen
from the fact that all contributions have the same form $\hat
\Omega_{IJ\,A}{}^B \, m^{I\Lambda}m^{J\Sigma} M_\Lambda \bar
M_\Sigma$ with the appropriate coefficients to make it vanish. The
only non--trivial rewriting is the one involving the square of the
gauginos, where using the identities of special geometry we find
that
\begin{equation}
h_{\Lambda\,i} h_{\Sigma\,\bar \jmath} g^{i\bar\jmath} =
-M_\Lambda \overline M_\Sigma + \frac12(Im {\cal
N})_{\Lambda\Sigma}
\end{equation}
and the second part is identically vanishing since it is
contracted with $\epsilon_{xyz}L^x_IL^y_J m^{\Lambda I} m^{\Sigma
J}$. At this point we are left with the mixed contributions which
give us a non--trivial constraint on the gauge group.

There are two different types of contributions to the Ward
identity (\ref{ward}) which are proportional to both the electric
and magnetic couplings and are triplets of $SU(2)$. The first one,
is given by $\frac{\rm i}{2} {\cal P}_{\Lambda AC} L_I^{CB} \,
m^{I\Sigma}\, L^\Lambda \overline M_\Sigma + c.c.$ and is common
to all the squares of the symplectic shifts. These again combine
with the appropriate coefficients to give zero. In addition, from
the gauginos one has a contribution which is an interference term
of the gauging of the vector multiplet isometries and the magnetic
gauging. This reads
\begin{equation}
g^{i\bar\jmath} W_{i AC} W_{\jmath}^{CB} \sim {\rm i} k^{i}_\Lambda \,
h_{\Sigma i}L_{IA}{}^C m^{I\Sigma} L^\Lambda + c.c.
\label{eq:extra}
\end{equation}
and it should vanish by itself.
The group--theoretic meaning of such equation becomes clear by using
the known identity of special geometry \cite{Andrianopoli:1997cm}
\begin{equation}
 k^i_\Lambda f_i^\Gamma = f^{\Gamma}_{\Lambda\Pi} \overline
 L^{\Pi}  + {\rm i} {\cal P}_\Lambda^0 \, V\,.
\label{eq:ide}
\end{equation}
Once this equation is used in (\ref{eq:extra}) the constraint becomes
\begin{equation}
f_{\Lambda\Sigma}^{\Delta} \, ({\rm Im} {\cal N})_{\Delta\Pi} \,
m^{I\Pi} = f_{\Lambda\Sigma}^\Delta  \tilde{m}^{I}_\Delta= 0\,,
\label{eq:contr}
\end{equation}
where
\begin{equation}
\tilde{m}^{I}_\Delta\equiv ({\rm Im} {\cal N})_{\Delta\Pi} \, m^{I\Pi} \,.
\end{equation}
Note that this equation means that $\tilde{m}^{I}_\Delta$ are the
coordinates of a Lie algebra element of the gauge group commuting
with all the generators, in other words a non trivial element of
the center ${\mathbb Z}({\mathbb G})$. Such a constraint follows
only for non--abelian gaugings because that is the only case when
one is forced to gauge explicitly the isometries of the vector
scalar manifold introducing the vector manifold Killing vectors
$k^i_\Lambda$. In our case, this condition is certainly satisfied
since we assumed from the beginning that the isometries that can
be gauged electrically are only those that commute with the
translational isometries of the dualized scalars. As it should be
obvious at this stage, the ``magnetic Killing vectors''
$m^{I\Lambda}$ are precisely associated to such translational
symmetries.

Once one has verified that the Ward identity (\ref{ward}) of
supersymmetry is satisfied for the $SU(2)$ Lie Algebra valued
pieces one can finally read the potential
\begin{equation}
\begin{array}{rcl}
{\cal V} &=& g^2\, \left\{ g_{i\overline{\jmath}}k^i_\Lambda k^{\overline{\jmath}}_\Sigma
\overline{L}^\Lambda L^\Sigma
+ 4 (g_{uv}+h_{Iu}M^{IJ}h_{Jv})k^u_\Lambda
k^v_\Sigma \, \overline{L}^\Lambda L^\Sigma
+ \left(U^{\Lambda\Sigma}
-
3 \overline{L}^\Lambda L^\Sigma\right) {\cal P}^x_\Lambda {\cal
P}^x_\Sigma \right\}\\[3mm]
&+& g\,\left[g^{i\bar\jmath} \left(f_i^\Lambda h_{\bar\jmath\Sigma} +
f_{\bar\jmath}^\Lambda h_{i\Sigma}\right)
- 3\left( \overline{M}_\Lambda L^\Sigma+ M_\Lambda
\overline{L}^\Sigma\right)\right]
{\cal P}^x_\Sigma {\cal
Q}^{x\Lambda}\\[3mm]
&-&2 g\,
h_{uI}\left(\overline{L}^\Lambda M_\Sigma
+ L^\Lambda\overline{M}_\Sigma\right) k^u_\Lambda\, m^{I\Sigma}\\[3mm]
&+& M_{IJ}
m^{I\Lambda}m^{J\Sigma}M_\Lambda\overline{M}_\Sigma
+\left(g^{i\bar\jmath} h_{i\Lambda} h_{\bar\jmath\Sigma}
- 3 \overline{M}_\Lambda M_\Sigma\right)
{\cal Q}^{x\Lambda} {\cal Q}^{x\Sigma}\,.
\end{array}
\label{eq:potential}
\end{equation}
In the absence of ``magnetic charges'' only the first line of 
(\ref{eq:potential}) is different from zero and corresponds to the 
standard ${\cal N} =2$ potential where $g_{uv}+h_{Iu}M^{IJ}h_{Jv}$ has 
to be identified with the metric of the quaternionic--K\"ahler 
manifold of the hypermultiplet sector.

It should be noted that the full potential (\ref{eq:potential}) is
symplectic invariant, being the square of symplectic invariant
quantities, except for the first term given by the square of the
Killing vectors of the special K\"ahler geometry. One can rewrite
${\cal V}$ by using (\ref{sympllvect}) and (\ref{eq:gradPQ}). In
doing so we find that the potential is a sum of
four distinct pieces
\begin{equation}
{\cal V}  =g^2\,  g_{i\overline{\jmath}}k^i_\Lambda
k^{\overline{\jmath}}_\Sigma \overline{L}^\Lambda L^\Sigma +
({\cal T}^x)^T {\cal M}_{S} {\cal T}^x +4\, \left[({\cal
Z}^\alpha_A)^T {\cal M}_N  \, {\cal Z}^A_\alpha -\, ({\cal T}^x)^T
{\cal M}_{N} {\cal T}^x\right]\,. \label{eq:pot2}
\end{equation}
where the scalar product of the last three terms, coming from the
square of symplectic invariant products, have been rewritten as
ordinary (orthogonal) matrix product. This implies that the
matrices ${\cal M}_{S} $ and ${\cal M}_{N}$
\begin{equation}
{\cal M}_S = -\frac{1}{2} \begin{pmatrix}
{ {\cal I}_{\Lambda\Sigma}+
({\cal R} {\cal I}^{-1}{\cal R})_{\Lambda\Sigma}&
-({\cal R}{\cal I}^{-1})_\Lambda{}^\Sigma
\cr
-({\cal I}^{-1}{\cal R})^\Lambda{}_{\Sigma}
&{\cal I}^{-1\,\Lambda\Sigma}
}
\end{pmatrix}
\label{eq:symplM}
\end{equation}
and
\begin{equation}
{\cal M}_N = \begin{pmatrix}
{
\overline{M}_\Lambda M_\Sigma &
-\overline{M}_\Lambda L^\Sigma
\cr
-\overline{L}^\Lambda M_\Sigma
&
\overline{L}^\Lambda L^\Sigma
}
\end{pmatrix}\,,
\label{eq:nonsymplM}
\end{equation}
where we introduced ${\cal I} \equiv Im{\cal N}$ and ${\cal R}
\equiv Re{\cal N}$ to make the notation compact, need not to be
symplectic (in spite of this ${\cal M}_{S}$ turns out to be
symplectic).

Since the term which explicitly breaks symplectic invariance
appears only when one gauges non--abelian groups any abelian
electric gauging leads to a symplectic invariant potential. It is
interesting to point out that precisely this type of gauged
supergravities with tensors appear naturally in compactifications
of type II string theories in the presence of non--trivial fluxes
for the Ramond--Ramond and Neveu--Schwarz three--form fields.
For the special case of Calabi--Yau compactifications, though, we do
not simply find that $g_{i\overline{\jmath}}k^i_\Lambda
k^{\overline{\jmath}}_\Sigma \overline{L}^\Lambda L^\Sigma =0$ as
expected, but another important cancellation happens.
For abelian isometries the definition of the prepotentials (\ref{prepotente}) simplifies
\cite{Michelson:1997pn,D'Auria:2001kv,Dall'Agata:2001zh} to
\begin{equation}
{\cal P}_\Lambda^x = \omega^x_{\hat u} k^{\hat u}_{\Lambda}\,.
\end{equation}
Once this is projected on the reduced scalar manifold following from
Calabi--Yau compactifications and we take
into account the analogous equation (\ref{Qdef}), we obtain that
\begin{equation}
({\cal Z}^\alpha_A)^T {\cal M}_N  \, {\cal Z}^A_\alpha = ({\cal
T}^x)^T {\cal M}_{N} {\cal T}^x\,, \label{eq:rela}
\end{equation}
which eventually implies that the scalar potential becomes
positive--definite and explicitly symplectic invariant
\begin{equation}
{\cal V} =  ({\cal T}^x)^T  {\cal M}_{S} {\cal T}^x \,.
\label{eq:POTENTIAL}
\end{equation}
We stress here that this result follows only for Calabi--Yau
compactifications and a standard choice of the symplectic sections
$V$, that is one for which the prepotential function of Special
Geometry  exists. In the generic case one cannot conclude that
(\ref{eq:rela}) holds, though usually analogous cancellations
between positive and negative terms of the potential also appear
for gaugings of translational isometries, again leading to
positive semi--definite potentials
\cite{Andrianopoli:2002vq,Andrianopoli:2003jf,Angelantonj:2003zx}.

\medskip

Let us now establish the relation between our work and the
standard formulation of \cite{Andrianopoli:1997cm}. At the level
of the ungauged theory the two formulations are simply related by
a dualization procedure, that is a Legendre transformation. Once
the theory is deformed by the gauging, the dualization of the
tensors into scalars cannot be performed anymore. 
For Abelian gaugings, though, we have seen that the scalar potential exhibits
an explicit symplectic invariance and one can use this invariance to 
put the potential in the standard form of \cite{Andrianopoli:1997cm}. 
In order to do so, one would like to remove all the dependence on the 
``magnetic charges'' $m^{I\Lambda}$, i.e. set ${\cal Q}^x = 0$. 
Since ${\cal T}^x$ is in general a local function of the 
coordinates we cannot rotate it to a configuration where ${\cal Q}^x = 
0$ by a constant symplectic matrix.
For special cases, though, 
if the number of abelian factors is the same as the number of tensors, 
one can find that the symplectic vector ${\cal T}^x$ 
can be written as an overall function of the scalars multiplying a 
vector given by 
constant electric and magnetic charges so that ${\cal T}^x \sim \{e^I_\Lambda, 
m^{I\Lambda}\}$, where 
the extra $I$ index on the electric charges follows from assigning different 
charges to the different abelian factors gauged.
However, since the ``magnetic
charges" have an extra index $I$, it is impossible to put all of 
them to zero by a symplectic rotation unless  the $\{e^I_\Lambda,
m^{I\Lambda}\}$ are parallel vectors for all $I$'s, or in the case
where there is just one tensor, $I=1$. The standard formulation
can thus be retrieved only in such cases. For non Abelian gaugings
we cannot in any case reduce the theory to the standard
formulation with only electric charges, due to the
presence of the non--symplectic invariant part.

The case of Abelian gaugings with parallel charge vectors is
realized in Calabi--Yau compactifications with two electric and
two magnetic charges coming from the Ramond--Ramond and
Neveu--Schwarz 2--forms of the Type IIB theory. In this case the 
tadpole cancellation condition implies that 
$e^1_\Lambda m^{2\Lambda}-e^2_\Lambda
m^{1\Lambda}=0$ (local case) \cite{Taylor:1999ii}. Therefore we can choose a
symplectic rotation such that the magnetic charges are set to zero
and the symplectic vector (\ref{sympllvect}) contains  only
the electric prepotentials. 
This also implies that the potential (\ref{eq:potential}) becomes of the form given in
\cite{Andrianopoli:1997cm,D'Auria:2001kv} for Abelian gaugings of
the quaternionic manifold. 
In detail, it can be shown to be 
\begin{equation}
{\cal V} =  ({\cal T}^{\prime x})^T {\cal M}_{S} {\cal T}^{\prime
x} = -\frac12 {\cal I}^{-1\Lambda\Sigma} {\cal P}^{\prime
x}_{\Lambda} {\cal P}^{\prime x}_\Sigma\,, \label{eq:finpot}
\end{equation}
which is of the form presented in
\cite{Taylor:1999ii,Dall'Agata:2001zh,Louis:2002ny}. 
Note, however, that, as shown in reference \cite{Kachru},  for
orientifold Calabi--Yau compactifications, the parallelism
condition cannot be imposed if we want to obtain a zero potential
at the extrema.

We conclude that, except for the afore mentioned particular cases,
the deformed theory in the presence of tensor
multiplets cannot be simply related to the one without them.
This seems to indicate that for such gaugings we find new
genuine deformations containing tensor fields which are
inequivalent to the formulation presented in
\cite{Andrianopoli:1997cm}. It would be very interesting to find a
derivation of these new supergravities as consistent effective
low--energy formulations of string or M--theory, possibly in the
presence of fluxes.

\bigskip

\phantom{E}

\bigskip

\noindent
{\bf Acknowledgments}

\medskip

We would like to thank L. Andrianopoli, A. Ceresole, A. Lled\'o,
M. Trigiante and S. Ferrara for valuable discussions. This work is
supported in part by the European Community's Human Potential
Programme under contract HPRN--CT--2000--00131 Quantum Spacetime,
in which R. D'Auria, S. Vaul\'a and L. Sommovigo are associated to
Torino University.

\medskip

\appendix

\section{Examples}

In this appendix we give two examples of how to retrieve the
geometry of the scalar manifold after dualization in the simple
case of the quaternionic manifolds ${\rm SU}(2,1)/{\rm
SU}(2)\times {\rm U}(1)$ and ${\rm SO}(4,1)/{\rm SO}(4)$.
Furthermore in the third example we justify our assertion that
dualization of a compact coordinate gives generally a singular
Lagrangian; this is done for the $\sigma$--model ${\rm
SU}(1,1)/{\rm U}(1)$.

Let us first discuss the dualization procedure for two simple
quaternionic manifolds, namely ${\rm SU}(2,1)/{\rm U}(2)$ and
${\rm SO}(4,1)/{\rm SO}(4)$.

\medskip

\noindent
{\bf Example 1: ${\rm SU}(2,1)/{\rm U}(2)$}

In accordance to our discussion in the text
(\ref{decomp}), we decompose the algebra of  ${\rm SU}(2,1)$ where
$\mathfrak{h}=\mathfrak{su}(2)+\mathfrak{u}(1)$, $\mathbb{D}=H$,
$\mathbb{N}=\{G,\,T_1,\,T_2\}$, where $H$ is the non compact
Cartan generator of $\mathfrak{su}(2,1)$ and $\{G,\,T_1,\,T_2\}$
are the nilpotent generators corresponding to the solvable algebra
generating the coset manifold. The commutation relations of the
solvable algebra are:
\begin{eqnarray}
\, [H,\,G]=G\,, &&[H,\,T_i]=\frac{1}{2}\,T_i\,,\quad
i=1,2\,,\\
\, [T_1,\,T_2]=G\,,&& [G,\,T_i]=0\,,
\end{eqnarray}
We choose as maximal abelian ideal the set of generators
$\{G,\,T_1\}$. The coset representative will be defined as
\begin{equation}
L=e^{\sigma G}\ e^{\varphi_1 T_1}\ e^{\varphi_2 T_2}\ e^{\phi
H}\,,
\end{equation}
where the fields $\{\sigma,\,\varphi_1,\,\varphi_2,\,\phi\}$ are
the scalar fields of the $\sigma$--model, the subset
$\{\sigma,\,\varphi_1\}$ being associated to the Peccei--Quinn
symmetries.

Let us first compute the metric of the quaternionic manifold ${\rm
SU}(2,1)/{\rm SU}(2)\times {\rm U}(1)$; we have
\begin{equation}
\label{ellemeno} L^{-1}dL=H d\phi+ \left(e^{-\frac{\phi}{2}}T_1+
e^{-\phi}\varphi_2G \right)d\varphi_1\ +e^{-\frac{\phi}{2}}T_2
d\varphi_2+ e^{-\phi} G d\sigma\,.
\end{equation}

Note that the nilpotent generators $\{G,\,T_1,\,T_2\}$ have a non
compact and a compact part. If  we denote with an hat the non
compact part and normalize the traces to a Kronecker delta,
namely:
 \begin{equation}\label{norm}
Tr(\hat G
H)=Tr(\hat{T_i}G)=Tr(\hat{T_i}H)=0\,,\quad\quad\quad\quad
Tr(\hat{T_i^2})=Tr(\hat{g^2})=Tr(\hat{H^2})=1\,,
\end{equation}
the metric is easily computed and we find:
\begin{equation}\label{metric2}
ds^2 = {\rm Tr}
\left(L^{-1}dL_{G/H}\right)^2=e^{-\phi}\left(d\varphi^2_1
    +d\varphi^2_2\right)
    +d\phi^2 +e^{-2\phi}\left(d\sigma +\varphi_2\,
    d \varphi_1\right)^2\,.
\end{equation}
It is very easy at this point to recover the scalar manifolds of
the $\sigma$--models associated to dualization of the coordinates
$\sigma$ and $\varphi_1$. It is sufficient to kill the associated
generators $\{G,\,T_1\}$ in equation (\ref{ellemeno}), so that the
metric of the resulting manifold is obtained  simply by setting to
zero the coordinates $\sigma$ and $\varphi_1$ and their
differentials in equation (\ref{metric2}).
We thus obtain that the
metric of the $\sigma$-- model of the double tensor multiplet is
given in this case by:
\begin{equation}
\label{double} ds^2=e^{-\phi}d\varphi_2^2+d\phi^2\,,
\end{equation}
which is easily seen to correspond to the coset manifold
${\rm SO}(2,1)/{\rm SO}(2)$.

If we instead dualize just one coordinate,
say $\sigma$, by the same procedure we find that the $\sigma$--
model of the single tensor multiplet has the metric:
\begin{equation}
\label{single} ds^2=e^{-\phi}\left(d\varphi^2_1
+d\varphi^2_2\right) +d\phi^2\,,
\end{equation}
which corresponds to the manifold ${\rm SO}(3,1)/{\rm SO}(3)$.

\medskip

\noindent
{\bf Example 2: ${\rm SO}(4,1)/{\rm SO}(4)$}

For the manifold ${\rm SO}(4,1)/{\rm SO}(4)$ we proceed in an
analogous way. We decompose the algebra $\mathfrak{g} =
\mathfrak{so}(4,1)$ in terms of $\mathfrak{h}$,  ${\mathbb D} = H$
and ${\mathbb N} = \{T_1,T_2,T_3\}$, where again $H$ is a non
compact Cartan generator of $\mathfrak{so}(4,1)$ and ${\mathbb N}$
describe the nilpotent ones. Explicitly, the $\mathfrak{so}(4,1)$
algebra is generated by $T^{ij}$ with $i,j=0,\ldots,4$, satisfying
\begin{equation}
[T^{ij},T_{kl}] = -4\,\delta^{[i}{}_{[k} T^{j]}{}_{l]}\,,
\label{eq:commso41}
\end{equation}
where indices are lowered by $\eta = {\rm diag} \{-++++\}$.
A solvable decomposition follows then by defining
\begin{equation}
H = T^{04}\,, \qquad T_{i} = T^{i0} - T^{i4}\,, \quad i = 1,2,3\,.
\label{eq:HN}
\end{equation}
The commutation relations of the solvable generators are
\begin{equation}
[H,T_i] = T_i\,, \qquad [T_i,T_j] = 0\,,
\label{eq:commso4}
\end{equation}
for any $i$.
The coset representative is now chosen as
\begin{equation}
L = \left(\Pi_i \, e^{\varphi_i T_i} \right) e^{\rho H}\,,
\label{eq:cosso4}
\end{equation}
where the fields $\{\rho, \varphi_i\}$ are the scalar fields of the
$\sigma$--model and the $x_i$ are associated with translational
isometries.
Proceeding as in the previous example we obtain for the metric of
$SO(4,1)/SO(4)$
\begin{equation}
\label{metric3}
ds^2 = {\rm Tr} \left(L^{-1}dL_{G/H}\right)^2= d\rho^2 +
e^{2\rho}\left(d\varphi_1^2+
    d\varphi_2^2 + d\varphi_3^2\right)\,.
\end{equation}
It is now easy to see that the resulting manifolds after
dualization of any of the $\varphi_i$ scalars yields $SO(3,1)/SO(3)$ and that
by the same procedure one can also obtain a double tensor multiplet
with scalar manifold $SO(2,1)/SO(2)$ and even a triple--tensor
multiplet with scalar manifold $SO(1,1)$.
More in detail, after an appropriate identification of $\rho$ with
$\phi$, by dualizing $\varphi_3$ one obtains (\ref{single}) and by
dualizing both $\varphi_3$ and $\varphi_1$ one gets (\ref{double}).

We stress that although the metric for the double tensor multiplet is
always (\ref{double}) one can dualize the remaining scalar $\varphi_2$
only for the theory coming from $SO(4,1)/SO(4)$ and not for the
universal hypermultiplet.
This happens because the metric of the kinetic term of the double tensor multiplet
coming from the $SU(2,1)/U(2)$ manifold contains explicitly
$\varphi_2$ whereas the kinetic terms of the tensors for the theory
following from $SO(4,1)/SO(4)$ depend only on $\phi$.

\medskip

\noindent
{\bf Example 3: ${\rm SU}(1,1)/{\rm U}(1)$}

In this last example we give an example of dualization of a
compact coordinate giving rise to a singular Lagrangian, taking as
toy model ${\rm SU}(1,1)/{\rm U}(1)$.

Let us consider the $\sigma$--model with metric, on the unit disk,
\begin{equation}
\label{disco}
    ds^2 = \frac{dz d \bar z}{(1 - z \bar z)^2}\,,
\end{equation}
or, introducing polar coordinates,
\begin{equation}
\label{polar}
    ds^2 = \frac{1}{(1- \rho^2)^2} (d\rho^2 + \rho^2 d\theta^2)\,.
\end{equation}

Suppose we dualize the compact coordinate $\theta$. Introducing
Lagrangian multipliers $H_\mu$ we have the following
$\sigma$--model Lagrangian
\begin{equation}
\label{lag}
    \mathcal{L} = \frac{1}{(1-\rho^2)^2}\partial_\mu \rho \partial_\nu \rho
    g^{\mu\nu}+ a H_\mu H^\mu +H^\mu \partial_\mu \theta\,.
\end{equation}
Varying $\theta$  we find $\partial_\mu H^\mu = 0$, that is
$H_\mu$ is the dual of the 3--form. Varying $H_\mu$ one finds
$H_\mu = -\frac{1}{2a} \partial_\mu \theta$. Substituting in
(\ref{lag}) and comparing with (\ref{polar}) fixes the value of $a
$ to be $a=-\frac{(1-\rho^2)^2}{\rho^2}$. The dualized Lagrangian
is then found by using $\partial_\mu \theta = -2a H_\mu$ in
(\ref{lag}) and one finally obtains
\begin{equation}
\label{dual}
    \mathcal{L}^{Dual} = \frac{\partial_\mu \rho\partial^\mu \rho}{(1
-\rho^2)^2} + \frac{(1-\rho^2)^2}{\rho^2} H_\mu H^\mu\,.
\end{equation}

Equation (\ref{dual}) shows explicitly that the Lagrangian is
singular in $\rho =0$ that is at the origin of the coset manifold.
If we instead set $z= \frac{i -S}{i+S}$ so that the unit disk is
conformally mapped into the upper half plane the corresponding
metric takes the form
\begin{equation}
\label{cocco1}
ds^2 = \frac{dS d\bar S}{(Im S)^2}\,,
\end{equation}
where we set $S=ie^\phi +C$. The same procedure used before gives
now, as dualization of the corresponding $\sigma$--model, the
following dual Lagrangian
\begin{equation}
\label{cocco}
    \mathcal{L}^{Dual} = \partial_\mu \phi\partial^\mu \phi +
    \frac{1}{4} e^{2\phi} H_\mu H^\mu\,,
\end{equation}
where no singularity appears. Indeed our parametrization with the
$S$ field corresponds exactly to performing the dualization of the
non--compact generator of the solvable Lie algebra of ${\rm
SU}(1,1)/{\rm U}(1)$



\providecommand{\href}[2]{#2}

\end{document}